\documentclass[prl,aps,showpacs,twocolumn]{revtex4}

\usepackage{times}
\usepackage{graphicx}
\usepackage{float}
\usepackage{latexsym,amsmath,amssymb,bm,euscript}
\usepackage{color}
\usepackage{subfigure}
\usepackage{epstopdf}
\usepackage[colorlinks=true,linkcolor=blue,citecolor=blue]{hyperref}

\usepackage{appendix}

\begin{document}

\title{Identifying Non-Abelian Topological Order  through Minimal Entangled States}

\author{W. Zhu$^1$, S. S. Gong$^1$, F. D. M. Haldane$^2$, D. N. Sheng$^1$}
\affiliation{$^1$Department of Physics and Astronomy, California State University, Northridge, California 91330, USA}
\affiliation{$^2$Department of Physics, Princeton University, Princeton, NJ 08544, USA}

\begin{abstract}
The topological order is equivalent to the pattern of long-range quantum entanglements,
which cannot be measured by any local observable.
Here we perform an exact diagonalization study to establish the non-Abelian topological order
through entanglement entropy measurement.
We focus on the quasiparticle statistics of
the non-Abelian Moore-Read and Read-Rezayi states on the lattice boson models.
We identify multiple independent minimal entangled states (MESs) in the groundstate manifold on a torus.
The extracted modular $\mathcal{S}$ matrix from MESs faithfully demonstrates the
Majorana quasiparticle or Fibonacci quasiparticle statistics, including
the quasiparticle quantum dimensions and the fusion rules for such systems.
These findings support that MESs manifest the eigenstates of quasiparticles
for the non-Abelian topological states and encode the full information of the topological order.
\end{abstract}


\maketitle

\textit{Introduction.---}
One of the most striking phenomena in the fractional quantum Hall (FQH) system
is the emergent fractionalized quasiparticles obeying  Abelian \cite{Laughlin} or
non-Abelian \cite{Moore,Greiter,Read} braiding statistics.
Interchange of two Abelian quasiparticles leads to a
nontrivial phase acquired by their wavefunction, whereas
interchange of two non-Abelian quasiparticles results in an operation of matrix
to the  degenerating groundstate space and
the final state will depend on the order of operations being carried out.
The non-Abelian quasiparticles and its braiding statistics
are fundamentally important for understanding the topological
order and  also %%%%statistics is far more exciting, especially for
have  potential application in topologically fault-tolerant
quantum computation \cite{Kitaev2003,Sarma,Nayak}.
So far, such  quasiparticles  have not been definitely identified in nature. However, it is generally
believed that they  exist in the FQH systems at filling factor $\nu=5/2$ \cite{Willett} and $12/5$ \cite{Xia},
described by the Moore-Read (MR)  \cite{Moore,Morf,Rezayi2000}
and Read-Rezayi (RR) states \cite{Read,Rezayi2009}, respectively.
%Moreover, experimental advances in the construction and the control of ultracold atoms in optical lattices have opened up another possibility of realizing the MR and RR states \cite{Cooper2001,Cooper2013, YFWang2012}.
The topological band model for optical lattices with bosonic particles is another
promising platform to realize the non-Abelian topological states \cite{Cooper2001,Cooper2013,YFWang2012,Haldane1988,Kapit2010,Tang,Neupert,Sun,DNSheng2011,Bernevig2011,YFWang2011}.
In the MR and RR states, the quasiparticles satisfy the following characteristic fusion rules that specify how the quasiparticles combine and fuse into more than one type of quasiparticles \cite{Nayak}:
%The  statistics of quasi-particles in non-Abelian model states can be described by %reflected in
%the following non-trivial fusion rules that specify how the
%quasiparticles combine and fuse into more than one type of quasiparticles\cite{Nayak}:
\begin{subequations}
\begin{align}
MR: && \sigma\times \sigma = \openone + \psi, \label{fusiona} \\
RR: && \tau \times \tau = \openone + \tau  \label{fusionb}
\end{align}
\end{subequations}
where $\openone$ represents the identity particle, $\psi$ the fermion-type quasiparticle,
$\sigma$ the Majorana quasiparticle and $\tau$ the Fibonacci quasiparticle.
In general, the fusion rule of quasiparticles
is encoded in the modular $\mathcal{S}$ matrix through Verlinde formula \cite{Verlinde,Wen1990,SDong,ZHWang,Fendley,Bonderson,Ardonne,Bais,Wen2012}.
%$a\times b=\sum_{c}N^c_{ab}c$ where $N^c_{ab}=\sum_{m}\mathcal{S}_{am}\mathcal{S}_{bm}\mathcal{S}^*_{mc}/\mathcal{S}_{1m}$.
Moreover, $\mathcal{S}$ ($\mathcal{S}_{i1}=d_i/\mathcal{D}$)
also determines the quasiparticle's individual quantum dimension ($d_i$) and total quantum dimension ($\mathcal{D}=\sqrt{\sum_i d_i^2}$)
\cite{Kitaev,Levin}.
Therefore, the $\mathcal{S}$ matrix plays the central
role in identifying topological order
and corresponding quasiparticle statistics \cite{Bonderson,Ardonne,Bais,Wen2012}.

While the Berry phase and non-Abelian information of quasiparticles  moving
adiabatically around each other have been studied
numerically \cite{Jain,Simon2003,Simon2009,Haldane2009,Lahtinen,Bloukbasi,Kapit2012},
the full modular $\mathcal{S}$ matrix and the corresponding fusion rules for the
microscopic models hosting the non-Abelian topological states have been lacking,
due to  the computational difficulty of directly dealing with and
distinguishing different quasiparticles.
Recently, there is growing interest on characterizing topological order through the quantum entanglement
information \cite{Kitaev,Levin,Haque,Isakov,HCJiang,Haldane2008,Lauchli2010,YZhang2012,Vidal,Tu,Pollmann}.
Among the recent progresses, the relationship between the entanglement measurement
and the modular matrix for topological nontrivial systems has been uncovered \cite{YZhang2012},
which may open a new avenue to this challenging issue.
The modular matrix and corresponding quasiparticle statistics
have been successfully extracted through the MESs for chiral spin liquid
and the Abelian FQH states \cite{YZhang2012,Vidal,WZhu},
which serve as direct evidences that the MES
is the eigenstate of the Wilson loop operators
with a definite type of quasiparticle \cite{YZhang2012,HCJiang,Vidal}.
For non-Abelian case, it is more challenging as the quasiparticles
usually have different quantum dimensions and topological entanglement entropies,
and it is unclear if the MESs still represent the quasiparticle eigenstates.
A recent variational quantum Monte Carlo calculation studied the quasiparticle statistics
for a projected  wavefunction \cite{YZhang2013},
however the obtained result does not accurately reproduce the known modular matrix
including each quasiparticle quantum dimension for the MR state;
thus it is of critical importance to
clarify if the MESs can lead to the accurate identification
of the  modular matrix and the corresponding topological order for microscopic non-Abelian
quantum states.

In this letter, we present an exact diagonalization (ED) study of
quasiparticle statistics of the possible MR and RR non-Abelian states
through extracting the modular $\mathcal{S}$ matrix for topological flat-band models \cite{Haldane1988,Kapit2010,Tang,Neupert,Sun,DNSheng2011,Bernevig2011}
with bosonic particles at filling numbers  $\nu=1$ and $\nu=3/2$.
We map out the entanglement entropy profile for
superposition states of the near degenerating groundstates and identify the global MES.
We find that all other MESs can be obtained
in the state space which is orthogonal to the global MES.
The obtained MESs form the orthogonal and complete basis states
for the modular transformation. We extract the modular $\mathcal{S}$ matrix
and establish the corresponding fusion rules, which unambiguously demonstrate the Majorana and Fibonacci quasiparticles
emerging in these systems. We demonstrate that the obtained modular matrix
as a topological invariant \cite{Bais} of the system remains to be universal in the whole topological phase
until a quantum phase transition takes place.

\begin{figure}[t]
 \begin{minipage}{0.49\linewidth}
  \includegraphics[width=1.5in]{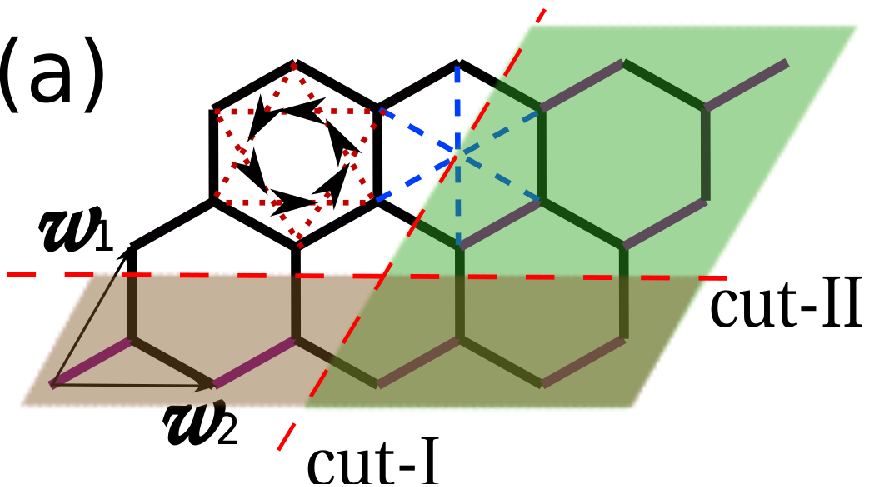}
 \end{minipage}
 \begin{minipage}{0.49\linewidth}
  \includegraphics[width=1.6in]{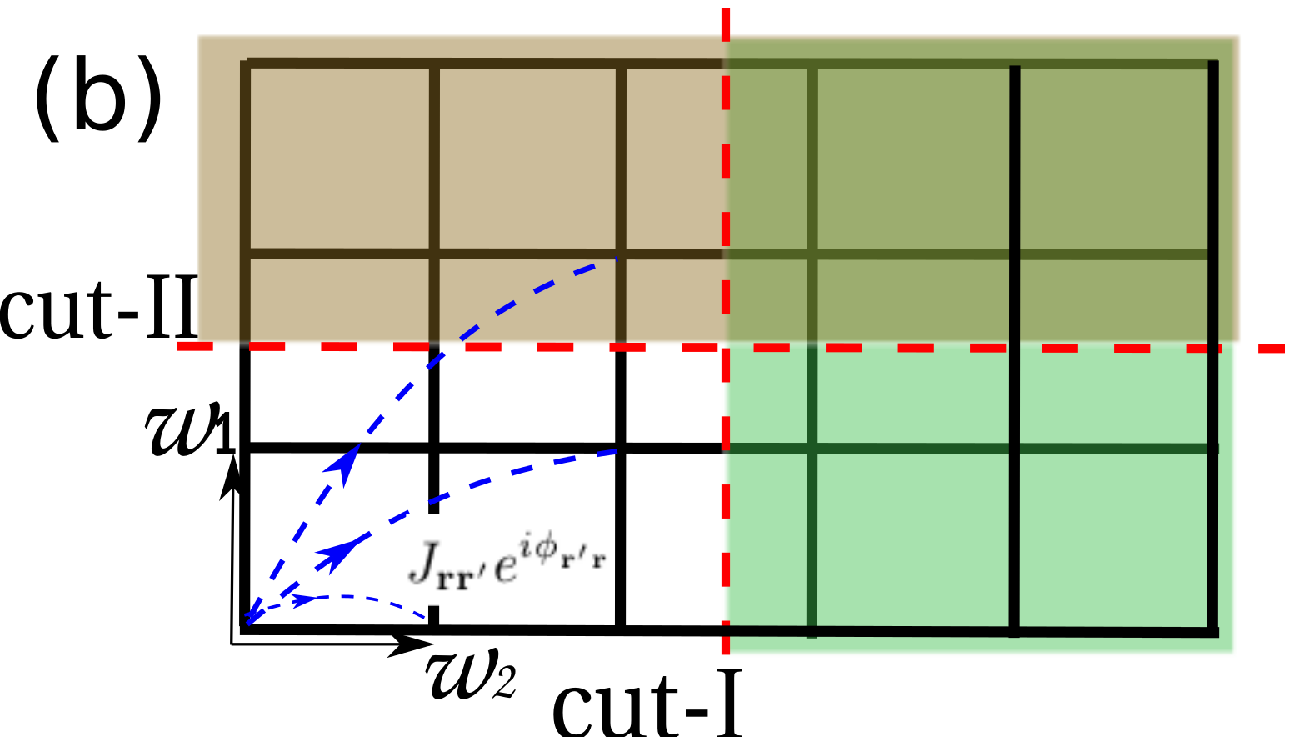}
 \end{minipage}
 \begin{minipage}{0.9\linewidth}
  \includegraphics[width=3.0in]{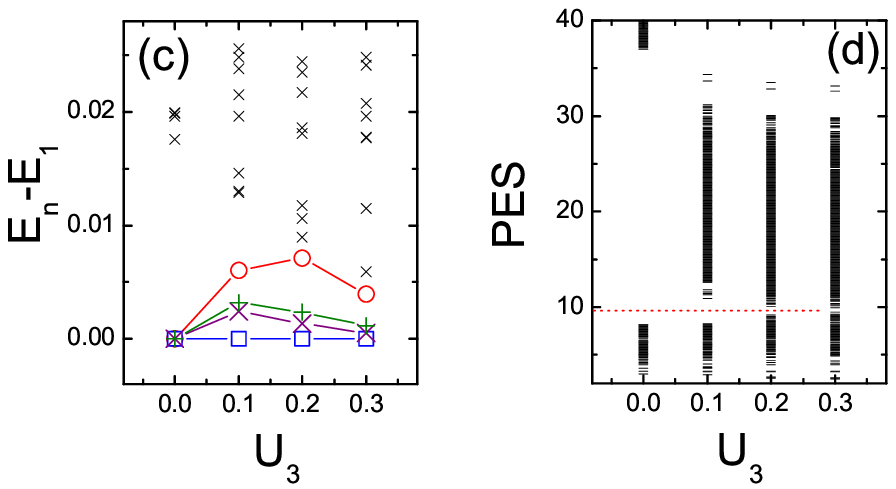}
 \end{minipage}
 \caption{(Color online)
 (a) Haldane model on HC lattice. The red dashed and blue dashed line represents the second NN hopping and third NN hopping, respectively.
 (b) SQ lattice with long-range hopping $J_{\mathbf{r}\mathbf{r}^{\prime}}e^{i\phi_{\mathbf{r}^{ \prime}\mathbf{r}}}$ as shown by blue dashed line.
 (c) Low-energy spectrum $E_n-E_1$ versus the $U_3$ ($U_2=0,U_4=\infty$) on a $4\times 4$ SQ lattice at $\nu=3/2$.
  Four lowest eigenvalues are labeled by blue square, purple cross, green cross and red circle.
 %(d) Low-energy spectra versus boundary phase $\theta_1$ at a fixed $\theta_2=0$ for $U_3=0.1$.
 (d) Particle Entanglement Spectrum (PES) for tracing out $4$ bosons. There are $298$ states below the PES gap
 (red dashed line) for $U_3<0.3$, in good agreement with the counting of quasihole excitations in RR state.
}\label{fig:lattice}
\end{figure}

%\textit{Hamiltonian.---}
We study the lattice boson model
with longer-range hoppings, which can be generally written as
\begin{eqnarray}\label{hamilton}
H=\sum_{\mathbf{r}\mathbf{r}^{\prime}}
\left[J_{\mathbf{r}\mathbf{r}^{\prime}}e^{i\phi_{\mathbf{r}^{ \prime}\mathbf{r}}}
b^{\dagger}_{\mathbf{r}^{ \prime}}b_{\mathbf{r}}
+\textrm{H.c.}\right]\nonumber + \sum_{n}\frac{U_n}{n!}\sum_{\mathbf{r}}(b^{\dagger}_{\mathbf{r}})^n (b_{\mathbf{r}})^n \\
\end{eqnarray}
where $b^{\dagger}_{\mathbf{r}} (b_{\mathbf{r}})$
creates (annihilates) a boson at site $\mathbf{r}=(x,y)$.
$U_n$ is an on-site N-body repulsive interaction.
Here we consider two representative lattice models:
the Haldane model on the honeycomb (HC) lattice \cite{Haldane1988,YFWang2012} and
topological flat band model on the square (SQ) lattice \cite{Kapit2010}.
On the HC lattice,
we include up to the third nearest-neighbor (NN) hopping
and a non-zero $\phi_{ij}$ on the second NN hopping only (the net flux is zero in one unit cell),
as shown in Fig.~\ref{fig:lattice}(a).
The NN hopping is set to be $J_{\mathbf{r}\mathbf{r}^{\prime}}=1$ and the other parameters
are defined the same as in Ref. \cite{YFWang2012}.
On the SQ lattice \cite{Kapit2010}, we select the phase factor $\phi_{\mathbf{r}^{ \prime}\mathbf{r}}$
corresponding to half flux quanta per plaquette.
The amplitude of hopping satisfies a particular gaussian form:
$J_{\mathbf{r}\mathbf{r}^{\prime}}=-
tG(\mathbf{r}-\mathbf{r}^{\prime})e^{-\frac{\pi}{4}|\mathbf{r}-\mathbf{r}^{\prime}|^2 }$,
where $G(\mathbf{r}-\mathbf{r}^{\prime})=(-1)^{(1+x-x')(1+y-y')}$
and $t=1$ as the energy scale here.
%so that the energy band can be flat even for small system sizes.
For both  models, we consider a finite size system with $N_x\times N_y$ unit cells,
the filling factor of lower band is $\nu=N_p/N_{s}$, where $N_p$ is boson number
and $N_{s}$ is number of single-particle states in the flat band.

We first obtain the low energy spectrum of the SQ model
at filling number $\nu=3/2$ and $\nu=1$ (see \cite{supplement}).
For $\nu=3/2$, we set $U_4=\infty$ so that
only three  bosons can go to the same lattice site,
which is effectively equivalent to a spin$-\frac{3}{2}$ system.
Due to much larger Hilbert space
than the conventional hardcore boson systems, the largest size we can deal with is limited
to  $4\times 4$ for $\nu=3/2$.
We find strong  numerical evidence of a possible
$\nu=3/2$ RR state and $\nu=1$ MR state (see \cite{supplement}) on the SQ lattice.
As shown in the Fig.~\ref{fig:lattice}(c-d), at smaller $U_3$ side,
we find robust fourfold degeneracy of groundstates on a torus
and the right counting rule of quasiparticle excitations for RR state \cite{Bernevig2012}.
The $\nu=1$ bosonic MR state on the HC lattice has also been identified in the previous study \cite{YFWang2012}.
Here our focus is to characterize the quasiparticle statistics of the above non-Abelian states
through calculating the modular $\mathcal{S}$ matrix.

To address the quasiparticle statistics, we first obtain the quasiparticle eigenstates
through determining the MESs on a torus \cite{YZhang2012}.
The entanglement entropy is defined as $S=-Tr \rho_A\log\rho_A$,
where the reduced density matrix $\rho_A$ is obtained
through partitioning the full system into two subsystems $A$ and $B$
and tracing out the subsystem $B$.
Here we consider two noncontractible bipartitions on torus geometry (cut-I and cut-II)
(Fig.~\ref{fig:lattice}(a-b)),
which is along the lattice vectors $w_1,w_2$, respectively.

\begin{figure}[t]
 \begin{minipage}{0.49\linewidth}
  \includegraphics[width=1.5in]{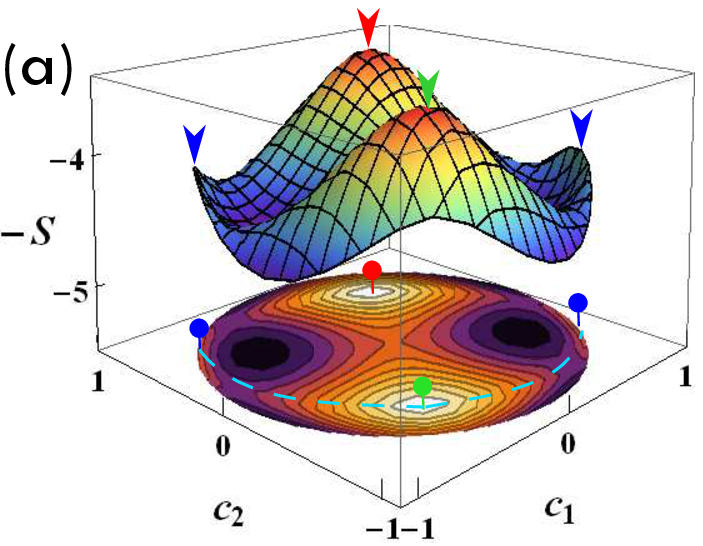}
 \end{minipage}
  \begin{minipage}{0.5\linewidth}
   \includegraphics[width=1.7in]{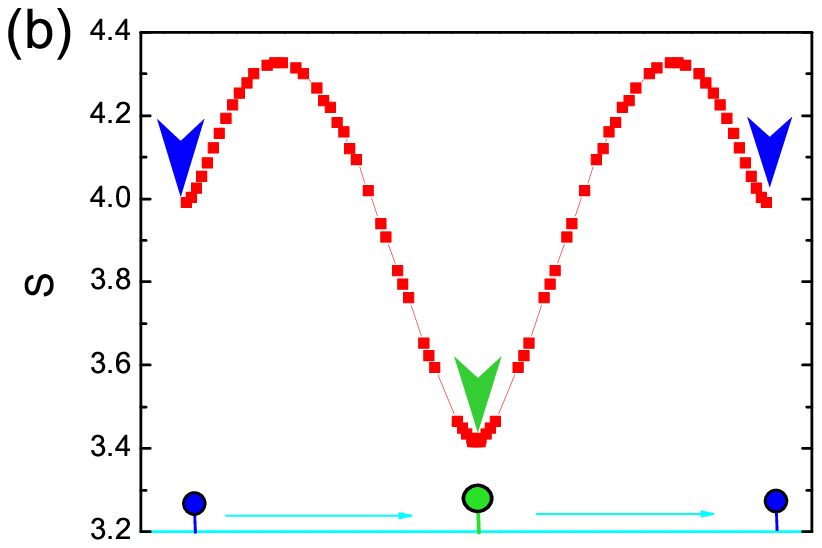}
  \end{minipage}
 \caption{(Color online)
  (a) Surface and contour plots of entanglement entropy ($-S$) of $|\Psi_{c_1,c_2,\phi_2,\phi_3}>$
  on $3\times 4$ HC lattice at $\nu=1$.
  We show entropy profile versus $c_1,c_2$ ($c_3=\sqrt{1-c_1^2-c_2^2}$)
  by setting optimized $\phi^o_2=1.26\pi$, $\phi^o_3=0.40\pi$.
  Three nearly orthogonal MESs are marked by red, green and blue arrows (dots) in surface (contour) plot.
  The cyan dashed line represents the states orthogonal to the first MES (red dot).
  (b) Entropy for the states along the cyan dashed line as shown in (a).
 All calculations are for partition along cut-I.
}\label{fig:HC:MR:MES}
\end{figure}

\textit{Moore-Read state at $\nu=1$.---}
We denote the three groundstates from ED calculation as $|\xi_j>$, (with $j=1,2,3$) \cite{YFWang2012}.
Now we form the general superposition states as,
\begin{equation*}
|\Psi_{(c_1,c_2,\phi_2,\phi_3)}>=c_1|\xi_1>+c_2e^{i\phi_2}|\xi_2>+c_3e^{i\phi_3}|\xi_3>
\end{equation*}
where $c_1,c_2,c_3$, $\phi_2$, $\phi_3$ are real superposition parameters.
For each state $|\Psi>$, we construct the reduced density
matrix and obtain the corresponding entanglement entropy.
We optimize values of $c_i\in[0,1]$ and $\phi_i\in[0,2\pi]$ to minimize the entanglement entropy.
In Fig.~\ref{fig:HC:MR:MES}(a),
we show the entropy profile at optimized parameters $(\phi^o_2,\phi^o_3)$ for MESs
on the HC lattice.
Here we draw the $-S$ in the surface and contour plots
so that the peaks in entropy show up clearly representing the minimums of $S$.
In Fig.~\ref{fig:HC:MR:MES}(a),
we find several peaks (entropy valleys) in $c_1-c_2$ space.
The first peak (red arrow) relates to the first (global) MES $|\Xi^{I}_1>$.
After determining $|\Xi^{I}_1>$, we search for the states with minimal entropy
in the state space orthogonal to $|\Xi^{I}_1>$ as shown in Fig.~\ref{fig:HC:MR:MES}(b).
The second and the third MESs are shown by green and blue arrows (the two states labeled by blue arrows are equivalent),
which are separately located in different entropy valleys as shown in Fig.\ref{fig:HC:MR:MES}(a).
We find that the first two MESs have almost the same entropy value,
indicating they are indeed topological equivalent with the same quantum dimension.
%%The entropy of each MES is listed in Table-\ref{table:entropy},
We calculate the entropy difference $\delta S$ between the third MES and the average of first two MES
to extract the information of the quantum dimension of the third quasiparticle.
%which is pretty big ($\delta S=0.576$) beyond a finite-size effect.
%We find the entropies corresponding to the three MESs are different from each other,
%which is also an indication of the different quantum dimensions of the quasi-particles.
As listed in Table-\ref{table:entropy},
nonzero $\delta S\approx 0.576$ implies the quantum dimension $d>1$ for the third quasiparticle state
\cite{Kitaev, Levin},
in agreement with the theoretical expectation for a MR state.
%Similarly, we study the SQ lattice and search for MESs in parameter space \cite{supplement};
%the entropies of the three nearly orthogonal MESs for both systems are  listed in Table-\ref{table:entropy}.
We also search for MESs for SQ lattice \cite{supplement} and find similar results as listed in Table-\ref{table:entropy}.

\begin{table}[t]
 \centering
 \caption{Entropy of MESs for $\nu=1$ and $\nu=3/2$.
 $S_i$ represents the entropy of $i-$th MES. For $\nu=1$, we use $\delta S=S_3-(S_2+S_1)/2$.
 For $\nu=3/2$, $\delta S=(S_3+S_4)/2-(S_1+S_2)/2$.
 $\delta S^*$ is analytic prediction \cite{Kitaev}.}
 \begin{tabular}{lcccccccc}
  \hline
  \hline
  $\nu$ & lattice size & $S_1$ & $S_2$ & $S_3$ & $S_4$ & $\delta S$ & $\delta S^*$ & $\delta S/\delta S^*$  \\
  \hline
  $1$ &HC $3\times 4$ & 3.416 & 3.416 & 3.991 & - & 0.576 & 0.693 & 0.831 \\ %& 3.204 \\
  $1$ &SQ $4\times 4$ & 2.546 & 2.924 & 3.357 & - & 0.622 & 0.693 & 0.897 \\ %& 2.076 \\
  $1$ &SQ $4\times 6$ & 2.528 & 2.971 & 3.443 & - & 0.694 & 0.693 & 1.001 \\ %& 2.076 \\
%  $1$ &SQ $4\times 4^*$ & 2.605 & 3.047 & 3.307 & - & 0.481 & 0.693 & 0.694 \\ %& 3.205 \\
  \hline
$\frac{3}{2}$ &SQ $4\times 4$& 3.168 & 3.732 & 4.157 & 4.417 & 0.837 & 0.961 & 0.871 \\ %& 3.693 \\
  \hline
  \hline
 \end{tabular}
 \label{table:entropy}
\end{table}

To extract the topological information of the quantum states from MESs,
we obtain  the overlap between the MESs for two noncontractible partition directions, which gives rise to
the modular matrix $\mathcal{S}=<\Xi^{II}|\Xi^{I}>$ \cite{YZhang2012}:
\begin{equation}\label{MR:HC:modularS}
\mathcal{S}\approx \frac{1}{2.033}
\left(\begin{array}{ccc}
        1.000 & 1.026 & 1.441 \\
        1.000 & 1.026 & -1.441 \\
        1.463 & -1.409 & 0.000
       \end{array}
     \right)
\end{equation}
for the HC lattice and we find very similar result for the SQ lattice \cite{supplement}.
Both results are quite close to the theoretical
result for MR state based on $SU(2)_2$ Chern-Simons theory \cite{SDong,ZHWang,Fendley}:
%\begin{equation}
$\mathcal{S}=\frac{1}{2}
\left(
       \begin{array}{ccc}
        1 & 1 & \sqrt{2} \\
        1 & 1 & -\sqrt{2} \\
        \sqrt{2} & -\sqrt{2} & 0
       \end{array}
     \right)$
%\end{equation}
, which determines the quasiparticle quantum dimension as:
$d_{\openone}=1$, $d_{\psi}=1$, $d_{\sigma}=\sqrt{2}$, $\mathcal{D}=2$ and non-trivial
fusion rule in Eq. (\ref{fusiona}).
From Eq. (\ref{MR:HC:modularS}),
we obtain $d^{HC}_{\sigma}\approx 1.463$, $\mathcal{D}^{HC}\approx 2.033$ for the HC lattice (and
$d^{SQ}_{\sigma}\approx 1.392$, $\mathcal{D}^{SQ}\approx 1.961$ for the SQ lattice \cite{supplement}).
Another striking point is that $\mathcal{S}_{33}\sim 0$,
which indicates that the two $\sigma$ quasiparticles
annihilate each other and fuse into
different quasiparticles.
To demonstrate this non-trivial behavior, we extract the fusion rule related to the third quasiparticle
from the numerical $\mathcal{S}$ matrix through Verlinde formula \cite{Verlinde}
$a\times b=\sum_{c}N^c_{ab}c$ where $N^c_{ab}=\sum_{m}\mathcal{S}_{am}\mathcal{S}_{bm}\mathcal{S}^*_{mc}/\mathcal{S}_{1m}$:
\begin{subequations}
\begin{align}
HC: && \sigma\times \sigma \approx 1.005\openone + 1.056\psi+0.096\sigma \label{MRa} \\
SQ: && \sigma\times \sigma \approx 0.951\openone + 0.959\psi+0.004\sigma  \label{MRb}
\end{align}
\end{subequations}
, which agrees excellently with the fusion rule of the MR state (Eq. \ref{fusiona}).
Both the $d_{\sigma}\approx\sqrt{2}$ and the characteristic fusion rule Eq. (\ref{MRa}-\ref{MRb})
unambiguously demonstrate the third quasiparticle $\sigma$ representing a Majorana fermion.
The fusion rule represents two ways to fuse two Majorana quasiparticles
therefore each pair of Majorana quasiparticles
can act as a qubit for quantum computation \cite{Nayak}.

\begin{figure}[b]
 \begin{minipage}{0.48\linewidth}
  \includegraphics[width=1.6in]{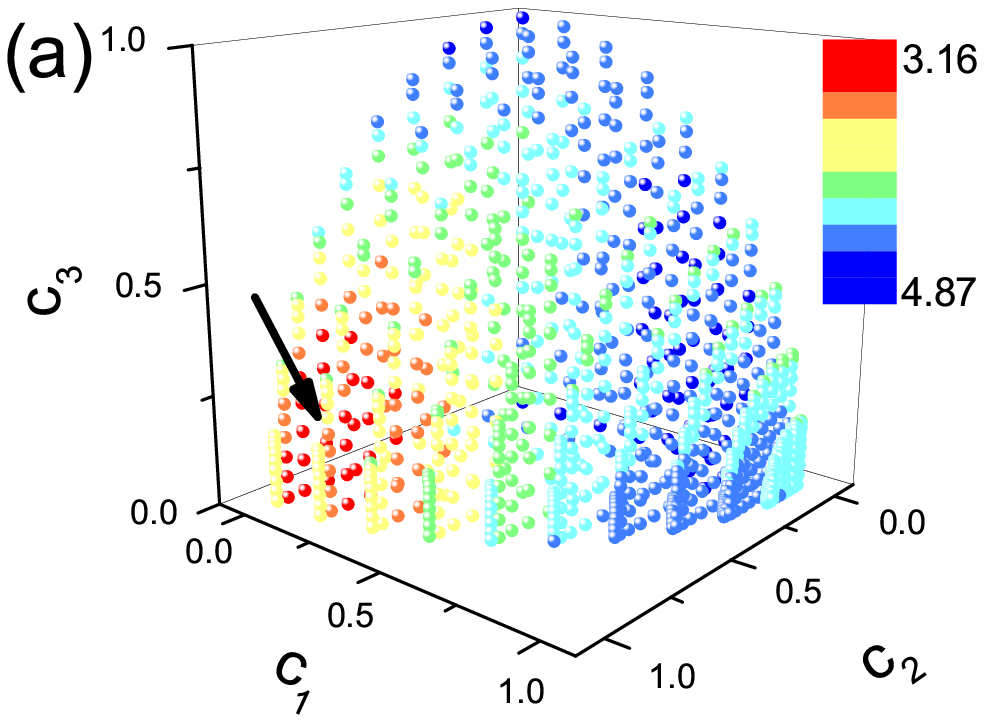}
 \end{minipage}
 \begin{minipage}{0.50\linewidth}
  \includegraphics[width=1.5in]{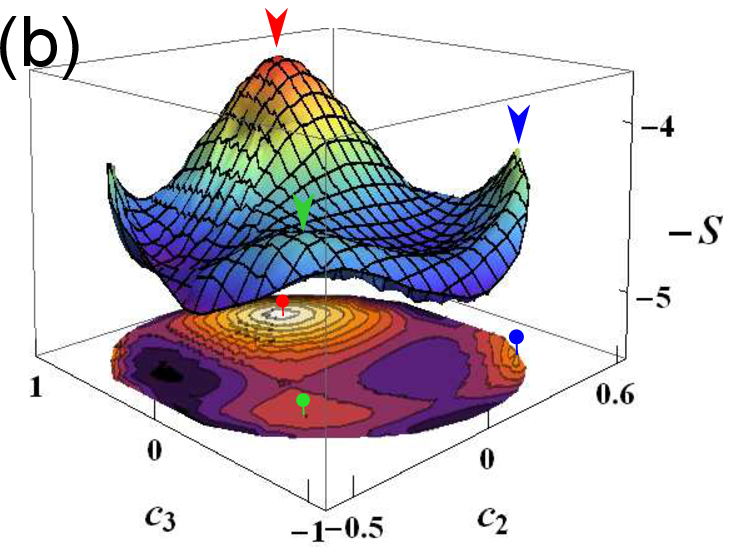}
 \end{minipage}
\caption{(Color online) (a) The entropy profile of wavefunction $|\Psi>$ at $\nu=3/2$
 in $c_1-c_2-c_3$ space by setting optimized $\phi^o_2=0.90\pi,\phi^o_3=0.46\pi,\phi^o_4=0.68\pi$.
 The color of dots represents the magnitude of entropy. The first MES indicated by black arrow.
 (b) The entropy profile versus $c_2-c_3$ in the space orthogonal to the first MES.
 The second, third and fourth MESs are labeled by red, green and blue arrows and dots, respectively.
 The calculation is for bipartition system along cut-I direction.
 }
\label{fig:SQ:RR:MES}
\end{figure}

\textit{Read-Rezayi State at $\nu=3/2$.---}
We turn to study the possible non-Abelian phase at $\nu=3/2$ and
detect the Fibonacci quasiparticle statistics emerging in this state.
Following the similar route for MR at $\nu=1$, we search for the MESs
in the space of the groundstate manifold
using the following general wavefunctions:
\begin{equation*}
  |\Psi> = c_1|\xi_1> +c_2e^{i\phi_2}|\xi_2> + c_3e^{i\phi_3}|\xi_3>+ c_4e^{i\phi_4}|\xi_4>
\end{equation*}
where $c_i$ and $\phi_i$ are the superposition parameters and $|\xi_j>$ ($j=1,2,3,4$)
are four groundstates from ED calculation.
We optimize the superposition parameters $c_i,\phi_i$
to minimize the entanglement entropy. In Fig.~\ref{fig:SQ:RR:MES}(a),
we show the global MES $|\Xi^{I}_1>$ in parameter space as labeled by the black arrow.
The other MESs $|\Xi^{I}_{i}>$ ($i=2,3,4$) are determined in the parameter space orthogonal
to $|\Xi^{I}_1>$ (Fig.~\ref{fig:SQ:RR:MES}(b)).
The entropies of the last two MESs are different from the lowest two MESs
as list in Table-\ref{table:entropy}, which is
consistent with the non-Abelian behavior of the quasiparticles.
However,  we also notice some finite size effect as all the four
entropies are different.

For the $\nu=3/2$ bosonic RR state, the edge
conformal field theory is captured by the $SU(2)_3$ Wess-Zumino-Witten model
\cite{SDong,Fendley,Bonderson,Ardonne},
whose modular matrix can be effectively described by
a non-Abelian $k=3$ $\mathbb{Z}_{k}$-parafermion part coupled by an Abelian semion part as:
$ \mathcal{S} = \mathcal{S}_{pf}\otimes\mathcal{S}_{U(1)}
              = \frac{1}{\sqrt{2+\phi}}
                                \left(
                                  \begin{array}{cc}
                                    1 & \phi \\
                                    \phi & -1 \\
                                  \end{array}
                                \right)\otimes
                  \frac{1}{\sqrt{2}}
                  \left(
                    \begin{array}{cc}
                      1 & 1 \\
                      1 & -1 \\
                    \end{array}
                  \right)$
, where $\phi=\frac{1+\sqrt{5}}{2}$ is golden ratio number.
As a comparison, we obtain the numerical modular matrix by calculating the overlap between
 the MESs along cut I and II:
%By locating MESs for both cuts,  we obtain the numerical modular matrix as
\begin{eqnarray}\label{RR:SQ:modularS}
\mathcal{S}&\approx&
\mathcal{S}_{pf}\otimes\mathcal{S}_{U(1)}
+10^{-2}\times
\left(
\begin{array}{cccc}
4.3 & -3.4 & -0.4 & 1.1 \\
-2.7 & -3.1 & -0.9 & -0.1 \\
2.1 & -0.8 & 2.4 & -0.7 \\
0.0 & -1.0 & -0.1 & -1.9 \\
\end{array}
\right) \nonumber\\
\end{eqnarray}
, which agrees with the analytic prediction (with a finite size correction of the order of  $10^{-2}$).
%The numerical modular matrix clearly represents the emergent of %the quantum dimension and the fusion rule of
%the Fibonacci quasi-particle in this bosonic RR $\nu=3/2$ phase.
The modular matrix $\mathcal{S}_{pf}$ signals Fibonacci quasiparticle
including the quantum dimension $d_{\tau}=\phi\approx 1.618$ and
the related fusion rule as shown in Eq. (\ref{fusionb}).
Two Fibonacci quasiparticles may fuse into an identity or a Fibonacci quasiparticle,
which is analogous to two $SU(2)$ spin-1/2's combining to either spin-1 or spin-0 total spin \cite{Feiguin2007}.
Using this property, Fibonacci quasiparticles is capable of
universal topological quantum computation \cite{Nayak}.

\begin{figure}[t]
\centering
 \begin{minipage}{0.49\linewidth}
  \includegraphics[width=1.5in]{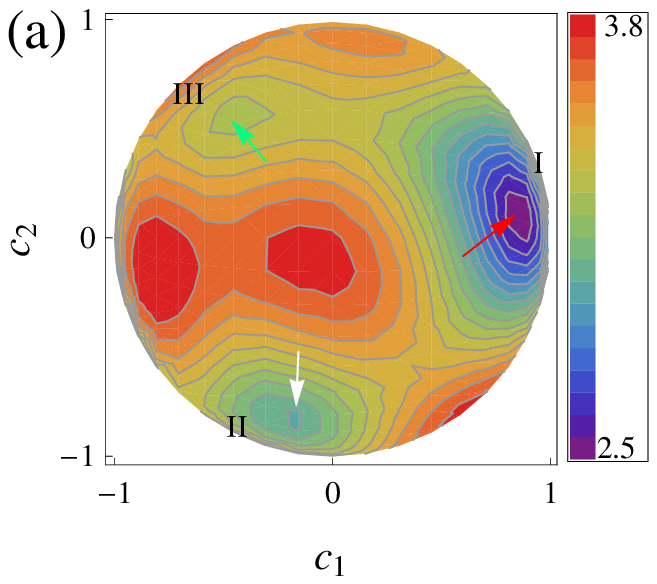}
 \end{minipage}
 \begin{minipage}{0.49\linewidth}
  \includegraphics[width=1.5in]{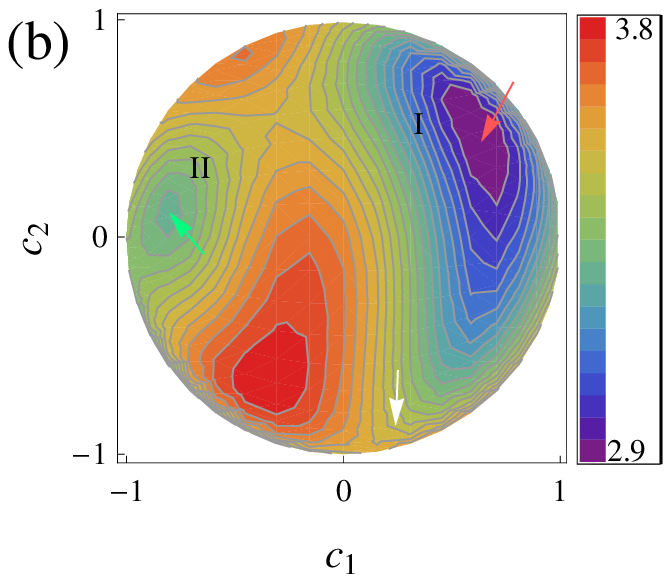}
 \end{minipage}
\caption{(Color online) The entropy of superposition state $|\Psi>$ in $c_1-c_2$ space ($c_3=\sqrt{1-c_1^2-c_2^2}$)
 for (a) $U_2=0.1$ and (b) $U_2=0.5$ by setting optimized $\phi_2,\phi_3$ at $\nu=1$ on the SQ lattice.
 }\label{fig:SQ:MR:QPT}
\end{figure}

\textit{Quantum Phase Transition.---}
By tuning the  interaction $U_n$, we
can drive  a quantum phase transition from the non-Abelian state to
other quantum phases\cite{YFWang2012,supplement}.
%PES also supports the topological phase vanishes when $U_2$ (or $U_3$) large enough.
A natural question is how the MESs and related modular $\mathcal{S}$ matrix evolve
around  the quantum phase transition region.
Here we study the SQ lattice model  at $\nu=1$ as example,
in which the quantum phase transition occurs around  $0.3<U^c_2< 0.4$ \cite{supplement}.
In the MR phase ($U_2=0.1$),
we find that the entropy profile of superposition state
has three valleys labeled by I, II and III,
as shown in Fig. \ref{fig:SQ:MR:QPT}(a).
The three orthogonal MESs are located in the above three valleys, respectively.
The resulting modular matrix remains  close to the theoretical one for the MR state:
$\mathcal{S}=
\frac{1}{1.965}
\left(
       \begin{array}{ccc}
        1.000 & 1.041 & 1.316 \\
        1.006 & 0.888 & -1.448 \\
        1.334 & -1.440 & 0.028
       \end{array}
\right)$.
We have also checked  that the $\mathcal{S}$ faithfully represents
the quasiparticle information for the whole phase region at $U_2 \leq 0.3$.
After the quantum phase transition at $U_2=0.5>U^c_2$,
we can only find two entropy valleys in entropy map,
as labeled by I,II in Fig. \ref{fig:SQ:MR:QPT}(b),
which relates to the first two MESs.
The third possible MES state (labeled by white arrow)  determined by the orthogonality relation,
actually is not a local minimum.
We continue to use these MESs as basis states to obtain the modular $\mathcal{S}$ matrix:
$\mathcal{S}=
\frac{1}{3.731}
\left(
       \begin{array}{ccc}
        1.000 & 2.873 & 2.037 \\
        2.899 & 0.750 & 2.354 \\
        2.015 & 2.354 & 2.082
       \end{array}
\right)$,
which deviates significantly from the MR $\mathcal{S}$ matrix.
In particular, the quasiparticle fusion rule and statistics has changed with $\mathcal{S}_{33}$
deviating from zero,  %% and $\mathcal{S}_{23},\mathcal{S}_{32}$ changing their signs,
which demonstrates the disappearance of the MR phase.

\textit{Summary and discussion.---}
We have numerically studied the non-Abelian quasiparticle statistics in the lattice boson models
which manifest the MR and RR non-Abelian states at filling factor $\nu=1$ and $\nu=3/2$, respectively.
Our work provides the first convincing demonstration of quasiparticle
fusion rules and statistics in microscopic topological band models.
The obtained modular $\mathcal{S}$ matrix faithfully represents
Majorana and Fibonacci quasiparticle statistics
including the quantum dimension for each quasiparticle and the fusion rules,
which fully support that the each MES is the eigenstate with a definite type of quasiparticle.
We are currently developing a numerical  method in DMRG simulations
to target different MESs by projecting out the previously identified
lower MESs, which we believe will become a useful tool for detecting the full information
of the topological order through modular matrix simulation in larger interacting systems \cite{preparation}.

\textit{Acknowledgements.---}
This work is supported by the US
NSF under grants DMR-0906816
(WZ,SSG), and by the Princeton MRSEC Grant DMR-0819860
(FDMH),  and the Department Of Energy Office
of Basic Energy Sciences under Grant No. DE-FG02-
06ER46305 (DNS).  DNS also acknowledges the travel support by
the Princeton MRSEC.

\clearpage
%\appendixpage

\begin{appendices}

\section{Supplemental materials for: ``Identifying Non-Abelian Topological order through Minimal Entangled States''}

In this supplemental material, we study the quasiparticle statistics of the non-Abelian Moore-Read (MR) state
on the square (SQ) lattice at filling factor $\nu=1$.
First, we find convincing evidence that the phase at filling factor $\nu=1$
manifests the MR state. Then we search for the minimal entangled states (MESs)
in the groundstate manifold and determine the modular $\mathcal{S}$ matrix.

\begin{figure}[b]
 \begin{minipage}{0.95\linewidth}
  \includegraphics[width=3.3in]{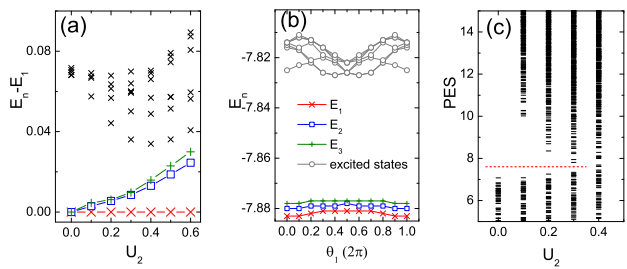}
 \end{minipage}
 \caption{(Color online)
 (a) Low-energy spectrum $E_n-E_1$ versus the $U_2$ ($U_3=\infty$) on a $4\times 4$ SQ lattice at $\nu=1$.
 Three lowest eigenvalues are labeled by blue square, red cross and green cross.
 (b) Low-energy spectra versus  boundary phase $\theta_1$ at a fixed $\theta_2=0$ for $U_2=0.1$.
 (c) Particle entanglement spectrum (PES) for tracing out $4$ bosons.
 There are $170$ states below the PES gap (red dashed line) for $U_2<0.4$,
 in good agreement with the counting of quasihole excitations in MR state.
}\label{fig:SQ:MR:ES}
\end{figure}

\textit{Low-energy Spectrum at $\nu=1$.---}
The Hamiltonian has been described in the main text and we select $U_3=\infty$ here
so that the system is effectively equivalent to a spin$-1$ system.
Low-energy spectrum versus different $U_2$ on the $4\times 4$ SQ lattice is shown in Fig.\ref{fig:SQ:MR:ES}(a).
When $U_2=0.0$, we find a groundstate manifold (GSM) with three exact degenerate lowest eigenstates.
%which separates from the higher eigenstates by a finite spectrum gap.
With increasing $U_2$, the exact degeneracy of GSM is destroyed and
two groundstates from the GSM will evolve into the excited spectrum.
We have also obtained numerical results from larger lattice size
($4\times 5$ and $4\times 6$) and have confirmed that the above picture is qualitatively correct.
%The Hilbert subspace of $4\times 6$ lattice have the dimensions of about $5\times 10^{7}$.
To confirm the robustness of this Fractional Quantum Hall (FQH) phase,
we also introduce two boundary phases $\theta_1$ and $\theta_2$
for generalized boundary conditions and
calculate the Chern number (Berry phase in units of $2\pi$) of GSM.
The three groundstates maintain their quasi-degeneracy and are well separated from the low-energy
excitation spectrum upon tuning the boundary phases (Fig.\ref{fig:SQ:MR:ES}(b)).
%, which indicates the robustness of this FQH phase.
Moreover, the three-fold GSM is found to share a total Chern number $C = 3$.
%which corresponds to a fractional quantized Hall conductance of $e^2/h$ per ground state.
%In order to investigate the nature of FQH at $\nu=1$,
Furthermore, we study the Particle entanglement spectrum (PES)
of the three-fold GSM (Fig.\ref{fig:SQ:MR:ES}(c)). PES
reveals a gap at $U_2<0.4$ and the number of states below this gap
exactly agrees with the number of quasihole excitations in a MR state
based on the generalized Pauli principle
that no more than two bosons occupy two consecutive orbitals \cite{Bernevig2012}.
Thus we reach the conclusion that the three-fold GSM at $\nu=1$ mimics
non-Abelian MR state at $U_2< 0.4$.
The PES gap disappears for $U_2\sim 0.4$ signaling the quantum phase transition
from the non-Abelian MR state to a topological trivial state \cite{YFWang2012}.

\begin{figure}[b]
 \begin{minipage}{0.49\linewidth}
  \includegraphics[width=1.6in]{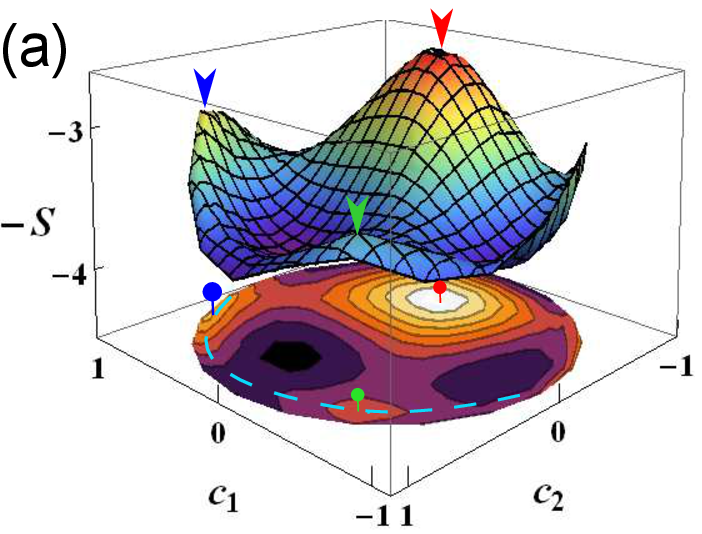}
 \end{minipage}
  \begin{minipage}{0.5\linewidth}
   \includegraphics[width=1.7in]{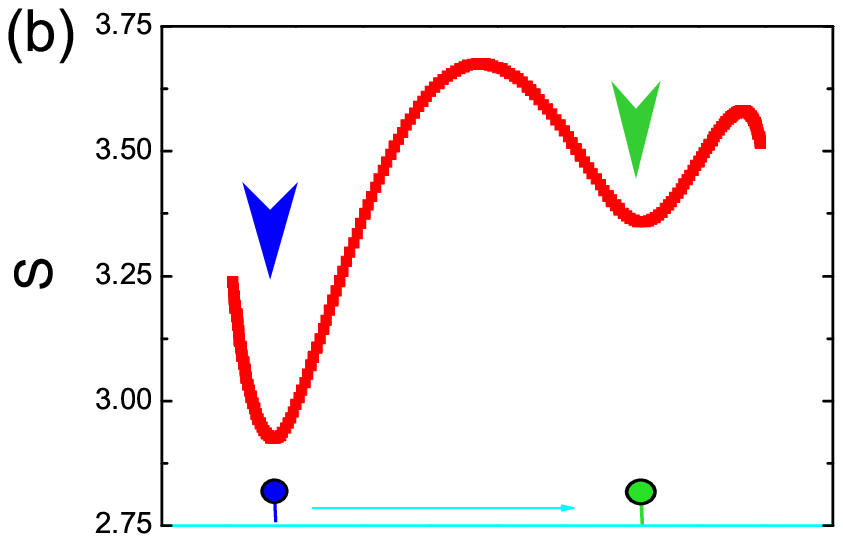}
  \end{minipage}
 \caption{(Color online)
  (a) Surface and contour plots of entanglement entropy ($-S$) of $|\Psi_{c_1,c_2,\phi_2,\phi_3}>$
  on $4\times 4$ SQ lattice at $\nu=1$.
  We show entropy profile versus $c_1,c_2$ ($c_3=\sqrt{1-c_1^2-c_2^2}$)
  by setting optimized $\phi^o_2=0.24\pi$, $\phi^o_3=0.52\pi$.
  Three nearly orthogonal MESs are marked by red, green and blue arrows (dots) in surface (contour) plot.
  The cyan dashed line represents the states orthogonal to the first MES (red dot).
  (b) Entropy for the states along the cyan dashed line as shown in (a).
 %(d) Entropy profile versus $\phi_2,\phi_3$ by setting $c_1=0.09$, $c_2=-0.49$, $c_3=0.87$.
 All calculations are for partition along cut-I.
}\label{fig:SQ:MR:MES}
\end{figure}

\textit{MESs and modular matrix at $\nu=1$.---}
We denote the three groundstates from ED calculation as $|\xi_j>$, (with $j=1,2,3$).
Now we form the general superposition state as,
\begin{equation*}
|\Psi_{(c_1,c_2,\phi_2,\phi_3)}>=c_1|\xi_1>+c_2e^{i\phi_2}|\xi_2>+c_3e^{i\phi_3}|\xi_3>
\end{equation*}
where $c_1,c_2,c_3$, $\phi_2$, $\phi_3$ are real superposition parameters.
For each state $|\Psi>$, we construct the reduced density
matrix and obtain the corresponding entanglement entropy.
We optimize values of $c_i\in[0,1]$ and $\phi_i\in[0,2\pi]$ to minimize the entanglement entropy.
In Fig.~\ref{fig:SQ:MR:MES}(a-b),
we show the entropy profile at optimized parameters $(\phi^o_2,\phi^o_3)$ for MESs.
In Fig.~\ref{fig:SQ:MR:MES}(a),
it is found several peaks (entropy valleys) in $c_1-c_2$ space and
the three nearly orthogonal MESs are determined as labeled by arrows.
In Fig.~\ref{fig:SQ:MR:MES}(b),
it is shown the entropies of states in the parameter space orthogonal to the first MES (red arrow).
The second and the third MESs are indeed located in the separated entropy valleys, respectively.
Here we find that the entropies corresponding to the three MESs are different from each other
(as list in Table-I in main text).
The entropy difference between the first MES and the second MES may result from the finite-size effect.
To identify the emergence of non-Abelian quasiparticles,
we calculate the modular matrix and obtain the quasiparticle statistics as below.

With the help of MESs, we obtain the modular matrix %$\mathcal{S}=<\Xi^{II}|\Xi^{I}>$:
\begin{equation}\label{MR:SQ:modularS}
\mathcal{S}\approx
\frac{1}{1.961}
\left(
       \begin{array}{ccc}
        1.000 & 1.025 & 1.373 \\
        0.929 & 0.920 & -1.431 \\
        1.392 & -1.376 & 0.037
       \end{array}
     \right)
\end{equation}
, which is close to the prediction of $SU(2)_2$ Chern-Simons theory:
%\begin{equation}
$\mathcal{S}=\frac{1}{2}
\left(
       \begin{array}{ccc}
        1 & 1 & \sqrt{2} \\
        1 & 1 & -\sqrt{2} \\
        \sqrt{2} & -\sqrt{2} & 0
       \end{array}
     \right)$.
%\end{equation}
We extract the three quasiparticle quantum dimension as
$d^{SQ}_{\openone}=1.000$, $d^{SQ}_{\psi}\approx 0.929$, $d^{SQ}_{\sigma}\approx 1.392$
and the total quantum dimension $\mathcal{D}^{SQ}\approx 1.961$.
%From the quantum dimension results,
%we find the third quasi-particle behaves non-Abelian quasiparticle ($d^{SQ}_{\sigma}\approx 1.392$),
%whereas the other two are Abelian type with quantum dimension approaching one.
%Moreover, the fusion rule of the third type quasiparticle $\sigma$ as
%$\sigma\times \sigma \approx 0.951\openone + 0.959\psi+0.004\sigma$ as shown in the main text,
%which is another evidence of its non-Abelian behavior.
Combined with the fusion rule of the third type of quasiparticle $\sigma$:
$\sigma\times \sigma \approx 0.951\openone + 0.959\psi+0.004\sigma$ (as shown in the main text),
we confirm that $\sigma$ behaves as the non-Abelian Majarona quasiparticle.

\end{appendices}

\end{document}